\documentclass[twocolumn,english,aps,prb,twocolum,superscriptaddress,bibnotes,amsmath,amssymb,floatfix]{revtex4-2}
\usepackage[colorlinks=true,citecolor=blue,linkcolor=magenta]{hyperref}

\usepackage[markup=nocolor, authormarkupposition=left]{changes}
\usepackage{soul}
\usepackage[utf8]{inputenc}
\usepackage[english]{babel}
\usepackage{amsmath,amsfonts,amssymb}
\usepackage[T1]{fontenc}
\usepackage{url}
\usepackage{changes}
\usepackage{amsmath}
\usepackage{amsfonts}
\usepackage{amssymb}
\usepackage{siunitx}
\usepackage{epstopdf}
\usepackage{graphicx}
\graphicspath{{./Figures/}}

\begin{document}

\title{Platicon microcomb generation using laser self-injection locking}

\author{Grigory Lihachev}
\affiliation{Institute of Physics, Swiss Federal Institute of Technology Lausanne (EPFL), CH-1015 Lausanne, Switzerland}

\author{Junqiu Liu}
\affiliation{Institute of Physics, Swiss Federal Institute of Technology Lausanne (EPFL), CH-1015 Lausanne, Switzerland}

\author{Wenle Weng}
\affiliation{Institute of Physics, Swiss Federal Institute of Technology Lausanne (EPFL), CH-1015 Lausanne, Switzerland}

\author{Lin Chang}
\affiliation{ECE Department, University of California Santa Barbara, Santa Barbara, CA 93106, USA}

\author{Joel Guo}
\affiliation{ECE Department, University of California Santa Barbara, Santa Barbara, CA 93106, USA}

\author{Jijun He}
\affiliation{Institute of Physics, Swiss Federal Institute of Technology Lausanne (EPFL), CH-1015 Lausanne, Switzerland}

\author{Rui Ning Wang}
\affiliation{Institute of Physics, Swiss Federal Institute of Technology Lausanne (EPFL), CH-1015 Lausanne, Switzerland}

\author{Miles H. Anderson}
\affiliation{Institute of Physics, Swiss Federal Institute of Technology Lausanne (EPFL), CH-1015 Lausanne, Switzerland}

\author{John E. Bowers}
\affiliation{ECE Department, University of California Santa Barbara, Santa Barbara, CA 93106, USA}

\author{Tobias J. Kippenberg}
\email[]{tobias.kippenberg@epfl.ch}
\affiliation{Institute of Physics, Swiss Federal Institute of Technology Lausanne (EPFL), CH-1015 Lausanne, Switzerland}

\maketitle
\noindent\textbf{The past decade has witnessed major advances in the development of photonic integrated, microresonator-based frequency combs (microcombs) \cite{Gaeta:19, Kippenberg:18}, that are coherent, broadband optical frequency combs with repetition rates in the millimeter-wave to microwave domain.
Integrated microcombs can be manufactured using wafer-scale process \cite{Moss:13}, and have been applied in numerous system-level applications \cite{Marin-Palomo:17, Corcoran:20, Obrzud:19, Suh:19, Trocha:18, Suh:18, Liang:15, Liu:20, Riemensberger:20, Spencer:18, Newman:19, Feldmann:21, Xu:21}.
Most of these advances are based on the discovery and harnessing of dissipative Kerr solitons (DKS) in optical microresonators with anomalous group velocity dispersion (GVD) \cite{Kippenberg:18, Herr:14}.
However, microcombs can also be generated with normal GVD using dissipative localized structures that are referred to as ``dark pulse'' \cite{Xue:15, Huang:15b, Xue:17b}, ``switching wave'' \cite{Parra-Rivas:16} or ``platicon'' \cite{Lobanov:15}.
Importantly, as most materials feature intrinsic normal GVD, in particular in the visible wavelength range, the requirement of dispersion engineering \cite{Luke:13, Okawachi:14}  is significantly relaxed for platicon generation.
Therefore, while DKS microcombs require particular designs and fabrication processes, platicon microcombs can be readily built using standard CMOS-compatible platforms such as thin-film (i.e. typical thickness below 300 nm) Si$_3$N$_4$ \cite{Munoz:19}.
Yet, laser self-injection locking \cite{Liang:15b, Liang:15} that has been recently used to create highly compact integrated DKS microcomb modules \cite{Raja:19, Shen:20, Voloshin:21} has not been demonstrated for platicons.
Here we report \cite{Lihachev:20} the first fully integrated platicon microcomb operating at a microwave-K-band repetition rate.
Using laser self-injection locking of a semiconductor distributed feedback laser edge-coupled to a high-$Q$ ($>10^7$) Si$_3$N$_4$ microresonator, platicons are electrically initiated and stably maintained, enabling a compact microcomb module without any complex control.
We further characterize the phase noise of the platicon repetition rate and the injection-locked pumping laser.
The observation of self-injection-locked platicons facilitates future wide adoption of microcombs as a build-in block in standard photonic integrated architectures via commercial foundry service.}

Optical frequency combs \cite{Udem:02, Cundiff:03} are coherent, broadband optical spectra consisting of equidistant grid of lines that have already revolutionized timing, spectroscopy and metrology \cite{Fortier:19, Diddams:20}.
In the last decade, microresonator-based Kerr frequency combs (``microcomb'') \cite{Kippenberg:11} have emerged as chip-scale optical frequency combs with repetition rates in the gigahertz to terahertz domain.
The operation of microcombs in the dissipative Kerr soliton (DKS) regime \cite{Herr:14, Kippenberg:18} has enabled fully coherent microcombs, and seeded applications such as coherent communication \cite{Marin-Palomo:17, Corcoran:20}, astronomical spectrometer calibration \cite{Obrzud:19, Suh:19}, ultrafast ranging \cite{Trocha:18, Suh:18}, low-noise microwave generation \cite{Liang:15, Liu:20}, massively parallel coherent LiDAR \cite{Riemensberger:20}, frequency synthesizers \cite{Spencer:18}, optical atomic clocks \cite{Newman:19}, and photonic convolutional neural network \cite{Feldmann:21, Xu:21}.
Nonlinear photonics materials for microcombs have vastly expanded, and now include integrated platforms such as silicon nitride (Si$_3$N$_4$) \cite{Moss:13, Xuan:16, Ji:17, Liu:21, Ye:19b, ElDirani:19}, AlN \cite{Jung:14, LiuX:18}, Hydex \cite{Wang:20, WangX:21}, LiNbO$_3$ \cite{He:19, Wang:19, Gong:20}, AlGaAs \cite{Pu:16, Chang:20}, GaP \cite{Wilson:20}, GaN \cite{Zheng:20} and SiC \cite{Lukin:20}, as well as air-cladded silica \cite{Yang:18, WuL:20}. 
Embracing the technological maturity of silicon photonics \cite{Thomson:16}, today fully integrated microcombs can be built on photonic integrated circuits (PIC) fabricated using CMOS-compatible materials and processes.
Silicon nitride has become the leading platform for integrated DKS microcombs \cite{Moss:13, Gaeta:19}.
Its inherent material properties exhibit 5 eV bandgap (leading to negligible two-photon absorption at telecommunication bands), high Kerr nonlinearity, low Raman and Brillouin nonlinearity \cite{Gyger:20}, and high power handling capability (more than 10 Watt continuous-wave (CW) laser power on-chip \cite{Brasch:15}) in tightly confining waveguides.
Key to its use for DKS microcombs is the ability to achieve ultralow propagating losses of 1 dB/m and microresonator quality factor $Q>10^7$, enabling DKS generation at milliwatt optical power level \cite{Xuan:16, Ji:17, Liu:21} that is already compatible with current III-V/Si lasers \cite{HuangD:19, Xiang:20, McKinzie:21, Xiang:21}.
Moreover, the significantly mitigated thermal effect simplifies DKS initiation without the requirements of fast or complex laser tuning schemes \cite{Brasch:16, Yi:16b, Stone:18, Zhou:19, Wildi:19}.

\begin{figure*}[t!]
\centering
\includegraphics[clip,scale=1]{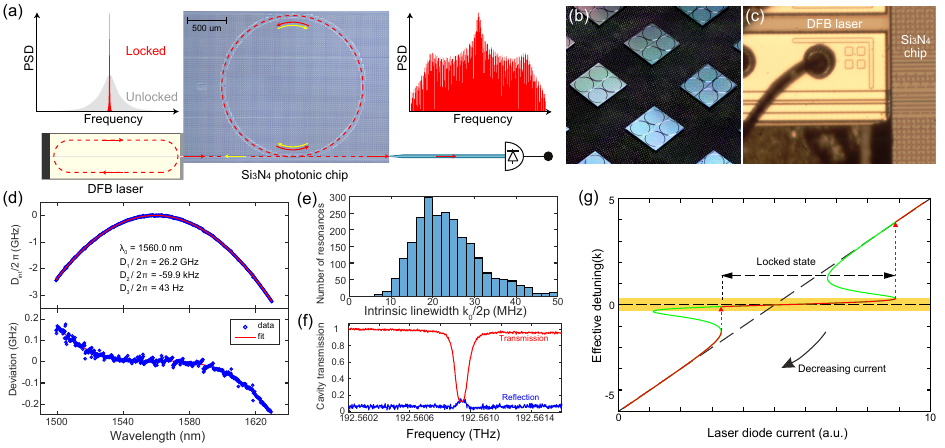}
\caption{
\textbf{Principle of platicon microcomb generation using laser self-injection locking.}
(a) Schematic of platicon generation using a DFB laser self-injection-locked to a Si$_3$N$_4$ photonic chip-based microresonator.
The bulk and surface Rayleigh scattering inside the Si$_3$N$_4$ microresonator induces back-scattered light that is sent back to the DFB laser and triggers laser self-injection locking.
This results in laser linewidth (i.e. frequency noise) reduction.
Further increasing the laser power can trigger platicon comb formation, as a result of the nonlinear interaction of counter-propagating cavity modes and modified laser dynamics due to the narrow-bandwidth Rayleigh back-scattering.
(b) Photograph of Si$_3$N$_4$ chips.
(c) Photograph showing the DFB laser edge-coupled to a Si$_3$N$_4$ waveguide.
(d) Top: Measured integrated dispersion of the Si$_3$N$_4$ microresonator.
This microresonator has an FSR of 26.2 GHz in the microwave K-band, and a normal GVD of $D_2/2\pi=-59.9$ kHz.
Bottom: The dispersion deviation from the $D_2$-dominant parabolic profile, to reveal the avoided mode crossings and higher-order dispersion terms.
(e) Histogram of 2414 fitted intrinsic resonance linewidth $\kappa_0/2\pi$.
The most probable value is $\kappa_0/2\pi=19$ MHz, corresponding to an intrinsic $Q_0=10^7$.
(f) Microresonator transmission (red) and reflection (blue) traces showing the resonance with 15$\%$ back-reflection.
(g) Simplified analytical estimation of laser frequency tuning curve based on the model of nonlinear self injection-locking (Ref. \cite{Voloshin:21})).
}
\label{fig1}
\end{figure*}

Despite the advancement mentioned, challenges remain.
It is well known that, for DKS generation, anomalous group velocity dispersion (GVD) is mandatory.
As Si$_3$N$_4$ has normal material GVD, dispersion engineering via waveguide geometry variation \cite{Luke:13, Okawachi:14} is required to access the anomalous-GVD regime, which necessitates film thickness above 600 nm for DKS generation at telecommunication bands.
When using low-pressure chemical vapor deposition (LPCVD) that offers Si$_3$N$_4$ films of superior quality (in terms of density, homogeneity, uniformity and stoichiometry), film cracks are often generated when the film thickness exceeds 500 nm.
To prevent crack formation, multiple schemes have been demonstrated, including using pre-structured substrates \cite{Nam:12, Pfeiffer:18b, Wu:20} and depositing Si$_3$N$_4$ in multiple cycles \cite{Gondarenko:09, Epping:15, ElDirani:18}.
However, these methods have not yet been incorporated into most commercial Si$_3$N$_4$ foundry processes, where LPCVD Si$_3$N$_4$ films with thickness below 400 nm are offered \cite{Munoz:19}.
Alternatively, plasma-enhanced chemical vapor deposition (PECVD) Si$_3$N$_4$ \cite{Chiles:18, Wang:18, Jin:20} and silicon-rich LPCVD nitride \cite{Ye:19, Liu:16b, Kruckel:17} are free from crack formation. 
However, these Si$_3$N$_4$ films have not yet obtained ultralow losses comparable to those achieved with LPCVD stoichiometric Si$_3$N$_4$.
Consequently, the wide adoption of DKS microcombs as a building block in standard PIC architectures via foundry service has not yet been possible.
Yet, coherent optical pulses (dissipative structures) can also be generated in optical microresonators with normal GVD .
This type of optical pulses is referred as ``dark pulse'' \cite{Xue:15, Huang:15b, Xue:17b}, ``switching wave'' \cite{Parra-Rivas:16} or ``platicon'' \cite{Lobanov:15}, and has been extensively studied in optical fibers \cite{Parra-Rivas:16} and observed in crystalline whispering-gallery-mode (WGM) resonators \cite{Liang:14}.
Compared with DKS, platicon has higher conversion efficiency from the CW pump to the pulse train \cite{Xue:17b}, ideal for applications such as microwave and millimeter-wave generation \cite{Liu:20, Wang:21}.
Platicons can be excited using pump amplitude modulation with the modulation frequency matching the multiples of the microresonator free spectra range (FSR), or by pulse-driving \cite{anderson2020zerodispersion}.
Conventional CW pumping scheme allows to initiate comb formation with a global normal GVD, by using the spatial mode coupling that leads to a local anomalous GVD near the pump \cite{Xue:15}.
Coupled-ring structures with thermal control of the location and coupling strength of avoided mode crossing \cite{KimB:19, Helgason:21} have also been employed.
Recently, switching dynamics of dark-pulse microcombs has been revealed and studied with ``hot'' cavity modulation spectroscopy technique \cite{Nazemosadat:21}.
In addition, dark-pulse microcombs have also been used for coherent data transmission \cite{Fulop:18}.

Here we demonstrate a compact, coherent, microwave-repetition-rate, platicon microcomb in a normal-GVD Si$_3$N$_4$ microresonator.
Figure \ref{fig1}(a) presents the operational scheme for platicon microcomb generation.
We use laser self-injection locking of a distributed-feedback (DFB) laser to a Si$_3$N$_4$ microresonator to seed the platicon.
Originally used to create narrow-linewidth lasers, in particular, compact lasers based on crystalline WGM resonators \cite{Liang:15b}, self-injection locking also offers routes to compact integrated DKS microcombs.
DKS formation has been reported in crystalline resonators \cite{Liang:15}, as well as in ultralow-loss Si$_3$N$_4$ chips using edge-coupled semiconductor lasers \cite{Raja:19, Shen:20, Voloshin:21}.
Optical spectrum of octave bandwidth has also been achieved \cite{Briles:20, Briles:21}.
Moreover, with external thermal and mechanical stabilization and using Si$_3$N$_4$ microresonators with negligible thermal effects \cite{Liu:21}, a new regime of DKS microcomb generation -- the ``turnkey'' operation \cite{Shen:20} -- has been revealed.

\begin{figure*}[t!]
\centering
\includegraphics[clip,scale=1]{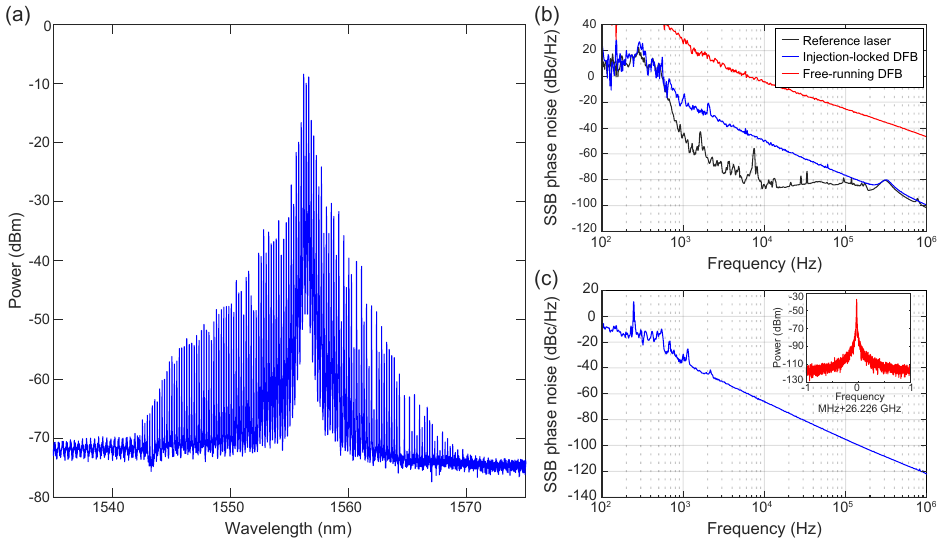}
\caption{
\textbf{Microwave-repetition rate platicon microcomb}.
(a) The optical spectrum of the coherent, single-FSR-spaced platicon microcomb with 26.2 GHz repetition rate.
(b) Comparison of single-sideband phase noise of free-running DFB (red) and self-injection-locked DFB (blue) beating against the reference ECDL laser (black).
The DFB laser is operated at 1556 nm.
The reference laser's phase noise (black) is measured against an ultrastable reference laser (MenloSystems ORNS) at 1552 nm.
(c) Single sideband phase noise of the beat signal among the platicon comb lines at 26.226 GHz, revealing the low noise nature.
Inset:  repetition rate beatnote signal.
}
\label{fig3}
\end{figure*}

In our experiment, a commercial DFB laser diode with a center wavelength of 1556 nm and output power up to 100 mW is used.
Through direct edge-coupling with an engineered inverse taper \cite{Liu:18} (cf Fig. \ref{fig1}(c)), around 20 mW optical power is coupled into the Si$_3$N$_4$ bus waveguide that is side-coupled to the Si$_3$N$_4$ microresonator.
The Si$_3$N$_4$ microresonator has a waveguide width of 2.00 $\mu$m, height around 530 nm, and a ring radius of 900 $\mu$m.
Such a waveguide geometry leads to a normal GVD.
We note that, different from thin-core Si$_3$N$_4$ waveguides \cite{Spencer:14, Gundavarapu:19, Jin:21, Puckett:21} that cannot be tightly bent due to the weak optical confinement, our waveguide geometry offers strong optical confinement allowing for the generation of platicon microcombs with repetition rates in the millimeter-wave domain \cite{Xue:15, Ye:19} to microwave domain (this work).

Frequency-comb-assisted diode laser spectroscopy \cite{DelHaye:09, Liu:16} is used to characterize the optical properties of the microresonator, including the dispersion profile and resonance linewidths.
This method comprises a mode-hop-free, widely tunable, external-cavity diode laser (ECDL) to acquire the optical transmission spectrum of the microresonator in the telecommunication band from 1500 to 1630 nm.
The transmission spectrum is frequency-calibrated using a commercial self-referenced fiber laser frequency comb \cite{DelHaye:09}, enabling accurate derivation of the linewidths of the microresonator resonances as well as the dispersion.
Here we only focus on the fundamental transverse-electric mode (TE$_{00}$) of the microresonator, as the edge-coupled DFB laser beam is TE-polarized.
Figure \ref{fig1}(d) top shows the measured integrated microresonator dispersion defined as $D_{\mathrm{int}}(\mu)=\omega_\mu-\omega_0-\mu D_1=D_2\mu^2/2+D_3\mu^3/6+D_4\mu^4/24+...$.
Here $\omega_\mu/2\pi$ is the frequency of the $\mu^{\mathrm{th}}$ resonance, $\omega_0/2\pi=192.62$ THz is the frequency of the pump resonance (corresponding to the laser wavelength of $\lambda_0=1556$ nm), $D_1/2\pi=26.2$ GHz is the microresonator FSR, $D_2/2\pi=-59.9$ kHz is the GVD parameter, $D_3/2\pi$ and $D_4/2\pi$ are higher-order dispersion terms.
To reveal the avoided mode crossings that are critical for platicon formation, the dispersion deviation from the $D_2$-dominant parabolic profile, defined as $[D_{\mathrm{int}}(\mu)-D_2\mu^2/2]/2\pi$, is shown in Fig. \ref{fig1}(d) bottom.
Figure \ref{fig1}(e) shows the histogram of 2414 fitted, intrinsic resonance linewidth $\kappa_0/2\pi$.
The microresonator is overcoupled in the telecommunication band, with the most probable $\kappa_0/2\pi=19$ MHz, corresponding to an intrinsic $Q_0=10^7$.

\begin{figure*}[t!]
\centering
\includegraphics[clip,scale=1]{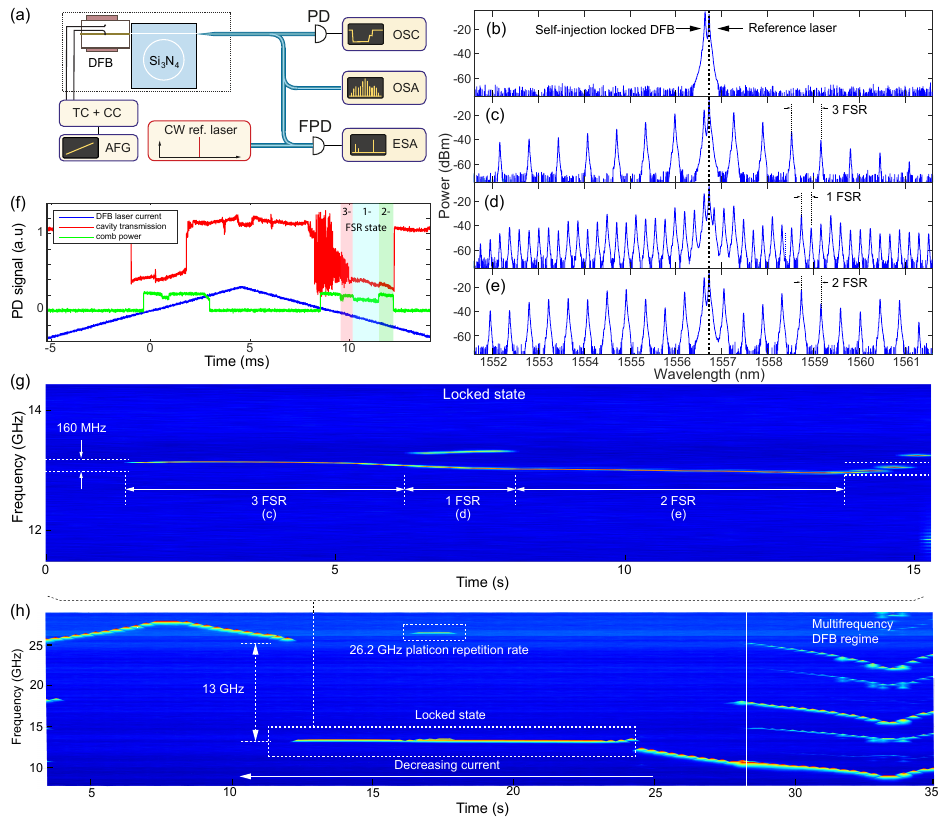}
\caption{
\textbf{Laser dynamics of self-injection-locked normal-GVD combs}.
(a) Experimental setup.
TC+CC: temperature and current controllers for the DFB laser.
AFG: arbitrary function generator.
PD, FPD: photodiode.
OSC: oscilloscope.
OSA: optical spectrum analyzer.
ESA: electrical spectrum analyzer.
(b-e) Optical spectra of different comb states upon backward tuning (decreasing laser current). 
Reference laser peak is at 1556.72 nm.
(f) Microresonator transmission trace (red) and generated comb light (yellow) upon the forward (increasing laser current) and backward scan of the DFB current (blue).
Shaded red, blue, green areas correspond to the formation of 3-, 1-, 2- FSR spaced comb.
(g-h) Beatnote spectroscopy of laser self-injection locking upon decreasing the DFB current, to reveal the laser dynamics. Signal at 13 GHz corresponds to a heterodyne beat of SIL laser and a reference laser, signal at 26.2 GHz is a platicon repetition signal.
}
\label{fig2}
\end{figure*}

\begin{figure*}[t!]
\centering
\includegraphics[clip,scale=1]{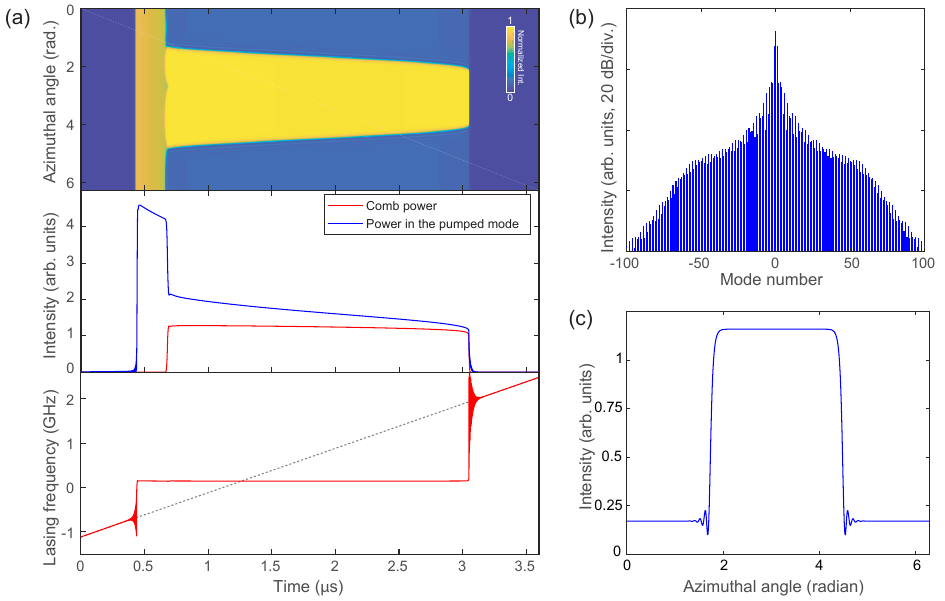}
\caption{
\textbf{Simulation of the microcomb generation with self-injection-locked laser pumping}.
(a) Dynamical evolutions of the microresonator intracavity intensity (upper panel), the intracavity comb power and the power in the pumped mode (middle panel), and the lasing frequency of the semiconductor laser (bottom panel) as the laser cavity resonance frequency is swept downwards across a microresonator resonance. The dashed line in the bottom panel indicates the lasing frequency trajectory if the optical feedback for self-injection locking is absent. (b) The platicon comb spectrum at the time of 2.1\,$\mu$s. (c) The temporal profile of the platicon corresponding to the microcomb spectrum in (b).
}
\label{fig4}
\end{figure*}


Laser self-injection locking is triggered via the Rayleigh backscattering in the microresonator, as outlined in Fig. \ref{fig1}(a).
The measured back-reflection from the resonance is 15$\%$ in power, corresponding to normalized forward-backward mode coupling $\beta=0.31$ (cf Fig. \ref{fig1}(f)).
First, we characterize the DFB laser noise by beating it with a reference laser and analyze the single-sideband (SSB) phase noise of the beat signal using a fast photodetector and an electrical spectrum analyzer (ESA).
The experimental setup is shown in Fig. \ref{fig2}(a).
An ECDL (Toptica CTL) PDH-locked to a reference Fabry-Pérot cavity of 30 kHz linewidth is used as the reference laser.
Figure \ref{fig3}(b) shows the measured SSB phase noise spectrum of the free-running DFB laser, the self-injection-locked DFB laser (the DFB laser current is 370 mA, same in both cases), and the reference laser.
At offsets above 300 kHz our measurement is limited by servo bumps from the PDH lock.
Compared to the free-running DFB laser, more than 30 dB phase noise reduction is observed in the self-injection-locked DFB, with Fourier offset frequency up to 1 MHz.
We also note that, at offsets frequency above 300 kHz, precise measurement of our DFB laser noise is limited by PDH servo bumps of our reference laser.
In the current case, further laser phase noise reduction can be obtained by optimizing the locking phase and fully packaging the DFB laser with the Si$_3$N$_4$ chip.
Furthermore, recent work \cite{Jin:21} using thin-core Si$_3$N$_4$ microresonators of high $Q$ and large mode volumes (to reduce the thermo-refractive noise \cite{Huang:19}) has achieved a laser phase noise performance comparable to that of fiber lasers.

By setting a proper DFB temperature and laser current, the coherent single-FSR-spacing platicon state can be directly accessed.
The platicon spectrum with 26.2 GHz line spacing (repetition rate) is shown in Fig. \ref{fig3}(a).
Spectral asymmetry and comb line power variation can be attributed to the influence of high-order dispersion terms \cite{anderson2020zerodispersion}.
Converted from 5 mW power of the CW DFB laser, the power of the platicon pulses collected in the output lensed fiber is measured to be around 1.5 mW.
Therefore the CW-to-platicon conversion efficiency is estimated to be 30\%, which shows nearly a two-order-of-magnitude improvement over the 0.4\% CW-to-soliton conversion efficiency in the previously reported bright DKS comb of 20 GHz repetition rate \cite{Liu:20}.
Upon photodetection of the platicon repetition rate using a fast InGaAs photodiode, we measure its phase noise spectrum with a phase-noise analyzer (PNA) using Welch's method from a time-sampling trace of the in-phase and quadrature components.
In the self-injection locking state without any external active feedback control to stabilize the platicon, the SSB phase noise of the platicon repetition rates around 26.2 GHz is shown in Fig. \ref{fig3}(c).

Next, we study in detail the platicon comb formation process.
With a carefully controlled locking phase (via varying the gap distance between the DFB laser and the Si$_3$N$_4$ chip), we tune the DFB laser current (thus the free-running laser frequency) over the microresonator resonance, and we observe multiple comb states switching with varying comb line spacings -- three-, one-, or two-FSR (see Fig. \ref{fig2}(c,d,e)).
Figure \ref{fig2}(f) shows the microresonator transmission spectrum of the DFB pump (red), the generated light (i.e. with the pump power filtered out) (green), upon triangular sweeping DFB laser current (blue).
The microresonator transmission spectrum shows characteristic vertical edges of the pumped resonance, marking the self-injection locking range.
We observe that tuning into coherent platicon states is feasible with both forward and backward scanning of the laser current, similar to the observation in anomalous-GVD microresonators as illustrated in Ref. \cite{Voloshin:21}.
However, we also note that the single-FSR-spaced (26.2 GHz) platicon state can only be achieved in the backward scan (i.e. decreasing laser frequency).
In order to obtain more insights into the comb formation dynamics, we employ a beatnote spectroscopy with an auxiliary frequency-stabilized laser and measure the nonlinear frequency tuning curve of self-injection-locked DFB.
Figure \ref{fig2}(h) shows the time-frequency spectrogram that the laser frequency scans across the resonance, corresponding to the backward tuning in Fig. \ref{fig2}(f).
When the DFB laser current is decreased linearly, the laser initially follows the linear tuning curve, and suddenly jumps to the locked state where the laser emission frequency remains nearly unchanged despite the laser current tuning.
As highlighted in Fig. \ref{fig2}(g), a maximum frequency change of 160 MHz ($\sim8\times\kappa_0/2\pi$) can be achieved across the full locking range.
Dashed arrows in Fig. \ref{fig2}(g) mark the detuning range for three-, single-, and two-FSR-spacing comb states.
Based upon the beatnote spectroscopy, the estimated locking range is $\sim13$~GHz.
Figure \ref{fig1}(g) shows the analytical estimation of the frequency tuning curve based on the nonlinear self-injection model developed in Ref. \cite{Voloshin:21} using the parameters in our experiment.

To analyze the dynamics of the laser-microcomb system, we develop a model where an injection-locked semiconductor laser is coupled to counter-propagating fields in a Kerr microresonator, which can be expressed as:
\begin{align}
\begin{split}\label{eq1}
\frac{dN}{dt} = & \frac{I}{eV} - \gamma N - aV(N-N_0) |E_\mathrm{L}|^2
\end{split}\\
\begin{split}\label{eq2}
\frac{dE_\mathrm{L}}{dt} = &  \left[\frac{(1 - i \alpha)}{2} (a V (N-N_0) - \frac{1}{\tau}) -i\Delta\omega \right]E_\mathrm{L} \\
& + \eta^{-1} \kappa_\mathrm{inj} e^{i \theta} A_\mathrm{ccw}
\end{split}\\
\begin{split}\label{eq3}
\frac{\partial{A_\mathrm{cw}}}{\partial t} = &  i \frac{D_{2}}{2} \frac{\partial^2{A_\mathrm{cw}}}{\partial \phi^2} + i g (|A_\mathrm{cw}|^2 + 2 |A_\mathrm{ccw}|^2) A_\mathrm{cw} \\
& -\left( {\frac{\kappa_\mathrm{r}}{2} + i(\omega_0 - \omega_\mathrm{L}) } \right){A_\mathrm{cw}} + i \kappa_\mathrm{sc} A_\mathrm{ccw} \\
& + \eta \kappa_\mathrm{inj} e^{i \theta} E_\mathrm{L}
\end{split}\\
\begin{split}\label{eq4}
\frac{dA_\mathrm{ccw}}{dt} = & i g \left(|A_\mathrm{ccw}|^2 + 2 \int_{0}^{2\pi}\frac{|A_\mathrm{cw}|^2}{2\pi} \,d\phi\right) A_\mathrm{ccw} \\
& -\left( {\frac{\kappa_\mathrm{r}}{2} + i(\omega_0 - \omega_\mathrm{L}) } \right){A_\mathrm{ccw}} + i \kappa_\mathrm{sc}^* A_\mathrm{cw}^0
\end{split}
\end{align}

Equations \ref{eq1} and \ref{eq2} are the semiconductor laser rate equations, where $E_\mathrm{L}$ represents the complex field, $I$ is the bias current, $N$ is the carrier density, $\alpha$ is the Henry factor, $a$ is the differential gain, $V$ is the laser active volume, $N_0$ is the carrier density at transparency, $\gamma$ is the carrier recombination rate, $\tau$ is the laser photon lifetime, $\Delta\omega$ is the offset frequency of the laser cavity resonance, and $e$ is the elementary charge.
Using $E_\mathrm{L}$ as the driving field, Equation \ref{eq3} is a modified Lugiato-Lefever equation (LLE) for the computation of the slowly varying field of the clockwise (cw) $A_\mathrm{cw}$, while Equation \ref{eq4} describes the dynamics of the counterclockwise (ccw) field $A_\mathrm{ccw}$. In these equations $\phi$ is the azimuthal angle, $D_2$ is the second-order dispersion coefficient, $\kappa_\mathrm{r}$ is the microresonator decay rate, $g$ is the single-photon-induced frequency shift due to the Kerr effect, and $\kappa_\mathrm{sc}$ is the backscattering-induced linear coupling rate between the cw and the ccw modes.
Since the power in the ccw mode is usually below the parametric four-wave-mixing threshold, here we treat the ccw field as $\phi$-independent, and only the field in the central mode of the microcomb ($A_\mathrm{cw}^0$) is scattered back to $A_\mathrm{ccw}$. As a result, while the cross phase modulation effect caused by the cw modes to the ccw modes takes the full power of the cw field profile into account with the integration $\int_{0}^{2\pi}\frac{|A_\mathrm{cw}|^2}{2\pi} \,d\phi$, the cross phase modulation caused by the ccw modes is simplified by using $|A_\mathrm{ccw}|^2$. The ccw field is injected back into the laser, with a feedback phase shift $\theta$ and a coupling rate between the microresonator mode and the laser mode that is denoted by $\kappa_\mathrm{inj} = \sqrt{\kappa_\mathrm{ex} \kappa_\mathrm{Lo}}$, where $\kappa_\mathrm{ex}$ is the external coupling rate of the microresonator, and $\kappa_\mathrm{Lo}$ is out-couple rate of the laser. In order to increase the computation efficiency, $|E_\mathrm{L}|^2$ is the photon density in the laser cavity for the semiconductor laser system, while $|A_\mathrm{cw}|^2$ and $|A_\mathrm{ccw}|^2$ are the intracavity photon numbers for the Kerr microresonator system.
To correctly relate the optical field profiles with different units in the laser (unit: m$^{-\frac{3}{2}}$) and the microresonator (unit: $1^{\frac{1}{2}}$), the parameter $\eta$ (unit: m$^{\frac{3}{2}}$) is introduced to the coupling between the laser and the Kerr microresonator.

Figure~\ref{fig4}(a) shows a typical microcomb generation process when the laser cavity resonance frequency is swept downwards over a microresonator mode. The values of the parameters used for this simulation is included in the Method section. As the laser cavity resonance approaches the microresonator resonance, the lasing frequency abruptly shifts to $\approx$150\,MHz at 0.44\,$\mu$s.
The lasing frequency is higher than the cold resonance frequency of the microresonator mode (i.\,e., 0~MHz) because the microresonator frequency is red-shifted due to Kerr effect when the laser is effectively coupled into the microresonator.
Shortly after, a platicon is formed by two opposite switching wave fronts, and the temporal duration of the platicon gradually decreases as the coupled-in laser power is reduced due to the increase of the laser-cavity-resonance-to-microresonator detuning. The platicon maintains stable existence until the laser exits from the self-injection-locked state, which shows that the whole frequency range of the injection locking state is over 2.5~GHz.
Figure~\ref{fig4}(b, c) show the microcomb spectrum at the time of 2.1\,$\mu$s and the corresponding temporal profile of the intracavity platicon, respectively.
Excellent agreement between the simulated comb spectrum and the experimental observation is achieved.
We note here that the results shown here are only one representative example, and that the successful microcomb generation and self-injection-locking of the laser can be obtained with varied feedback phases and lasing powers.

In summary, we demonstrated coherent platicon microcomb formation in an integrated, normal-GVD Si$_3$N$_4$ microresonator with a microwave-K-band repetition rate.
Employing laser self-injection locking, the platicon state is directly initiated via simply laser current tuning.
Different from previous works, our approach does not require complex coupled microrings and heaters that are required to control the avoided-mode crossing.
Moreover, our scheme could also be applied to the wavelength range, e.g. the visible wavelength, where microcomb generation is desired but anomalous dispersion engineering is infeasible.

\textit{Note added}:
Our preliminary result has been published in conference proceedings in Ref. \cite{Lihachev:20}.
During the preparation of this work, we note a work showing dark-pulse generation in thin-core Si$_3$N$_4$ microresonator \cite{Jin:21}.

\begin{footnotesize}
\noindent \textbf{Methods}:
The numerical simulation is carried out with the Runge-Kutta method, with the values of parameters $I=250$\,mA, $V=20\times10^{-17}$\,m$^3$, $\gamma=1\times10^9$\,Hz, $a=1\times10^4$\,Hz, $N_0=1\times10^{24}$, $\alpha=5$, $\tau=5\times10^{-12}$\,s, $\frac{\kappa_\mathrm{r}}{2\pi}=100$\,MHz, $\kappa_\mathrm{ex}/2\pi=50$\,MHz, $\kappa_\mathrm{sc}/2\pi=10$\,MHz, $\theta=\pi/2$, $\kappa_\mathrm{Lo}/2\pi=4.6$\,GHz, $\eta=2.6\times10^{-7}$\,m$^{1.5}$, $g=0.56$\,rad, $D_1/2\pi=26$\,GHz, and $D_2/2\pi=-60$\,kHz. Complex random noise is added into both the laser field and the microresonator intracavity fields to initiate the lasing and the platicon formation. The sweeping time in Fig.\,\ref{fig4} is 3.6\,$\mu$s in lab time, corresponding to 360 microresonator photon decay times. Such a long sweeping time eliminates transient instabilities after the platicon is created, allowing us to confirm the high stability and the coherence of the platicon structure.

\noindent \textbf{Acknowledgments}:
This work was supported by Contract HR0011-15-C-055 (DODOS) from the Defense Advanced Research Projects Agency (DARPA), Microsystems Technology Office (MTO),
by the Air Force Office of Scientific Research, Air Force Materiel Command, USAF under Award No. FA9550-19-1-0250,
by Swiss National Science Foundation under grant agreement No. 176563 (BRIDGE),
and by the European Union H2020 research and innovation programme under FET-Open grant agreement no. 863322 (TeraSlice).
The Si$_3$N$_4$ samples were fabricated in the Center of MicroNanoTechnology (CMi) at EPFL.
We thank Freedom Photonics for providing the DFB laser, and Dave Kinghorn for contribution in laser packaging.

\noindent \textbf{Data Availability Statement}:
The code and data used to produce the plots within this work will be released on the repository \texttt{Zenodo} upon publication of this preprint.

\end{footnotesize}

\bibliographystyle{apsrev4-1}
\bibliography{bibliography}
\end{document}